\documentclass[12pt,preprint]{aastex}
\usepackage{emulateapj5}

\def\fmag{\hbox{$.\!\!^m$}}

\newcommand{\mincir}{\raise
-2.truept\hbox{\rlap{\hbox{$\sim$}}\raise5.truept\hbox{$<$}\ }}
\newcommand{\magcir}{\raise
-2.truept\hbox{\rlap{\hbox{$\sim$}}\raise5.truept\hbox{$>$}\ }}

\shorttitle{cD galaxies}
\shortauthors{Tovmassian}

\begin{document}

\title{On the Formation and Evolution of cD Galaxies}

\author{Hrant M. Tovmassian}
\affil{Instituto Nacional de Astrof\'{\i}sica \'Optica y 
Electr\'onica, AP 51 y 216, 72000, Puebla, Pue, Mexico}
\email{e-mail: hrant@inaoep.mx}

\begin{abstract}
The cannibalism mechanism of formation of cD galaxies in clusters of Bautz-Morgan class I is analyzed. Dependences between absolute magnitude of cD galaxies, and of the clusters in which they reside (richness, redshifts) are considered. Evidences are presented in favor of formation of cD galaxies by cannibalism and their evolution.
\end{abstract}


\section{Introduction}

The formation mechanism of the brightest cluster galaxies (BCGs) is an important problem of modern astronomy (e.g. Lin \& Mohr 2004; von der Linden et al. 2007; Hansen et al. 2007; Garijo, Athanassoula, \& Garcia-G\'{o}mez 1997; Tutukov, Dryumov, \& Dryumova 2007). The definition of a BCG is based on the brightness contrast between it and the other cluster members. Some of BCGs are cD galaxies which have characteristic faint and extended halo. The physical properties of these unique objects are reviewed by Tonry (1987), Kormendy \& Djorgovski (1989) and Schombert (1992).

According to one of the proposed scenarios, BCGs are formed in the cluster cooling flows, when the gas density grows enough to cool and condense, forming stars in the cluster core (Silk 1976; Cowie \& Binney 1977; Fabian 1994). In this scenario there should be color gradients which are absent (Andreon et al. 1992). Another proposed possibility for formation of cDs involves tidal stripping by cluster galaxies which pass near the cluster center. The stripped material falls to the center of the potential well and may form the halo of the giant galaxy there (Gallagher \& Ostriker 1972; Richstone 1975, 1976). This theory can not explain, however, the difference between central dominant galaxies in clusters with and without a prominent halo. Garijo, Athanassoula, \& Garcia-G\'{o}mez (1997) mention in addition that the velocity dispersion of stars in cD halos is by three times smaller than the velocity dispersion of galaxies in the cluster, and so this theory has a difficulty in explaining why the tidally stripped material is slowed down as it builds up a cD halo.

The third hypotheses on formation of cDs is by rapid merging of galaxies during cluster collapse (e.g., Merritt 1983, 1984,1985: Dubinski 1998). However, as Merrit argues, the truncation of galaxy halos during cluster collapse should make time scales for dynamical friction longer than a Hubble time and thus "turn off" subsequent evolution in the cluster.

Note, that cD galaxies formed by the mentioned scenarios should be located exactly in the center of corresponding cluster and must have the mean radial velocity of the cluster galaxies. Meanwhile, some cDs are located at appreciably high distance from the geometrical center of the cluster and have peculiar velocities in relation to the mean cluster velocities of the order of 50-400 km s$^{-1}$ mentioned by many (e.g. Oegerle \& Hill 2001, and references therein). This fact poses problems for the mentioned mechanisms of formation of BCGs.

The most popular hypothesis is the galactic {\it cannibalism} (Hausman \& Ostriker 1978; Ostriker \& Trimain 1975; White 1976). Baier \& Schmidt (1992) showed that merging processes play an important role in the formation of cDs. According to this hypothesis, massive cluster galaxies gradually sink to the bottom of the potential well of a cluster and merge with each other (Searle, Sargent \& Bagnuolo 1973; Dressler 1980; Barnes 1989). Cannibalism may also be responsible for the further growth of giant galaxies formed by cooling flows or cluster collaps. 

In this paper I consider correlations between different parameters of BCGs and of the clusters in which they reside. Assuming the cannibalism scenario for formation of BCGs I present evidences on their evolution. In the study the emphasis is given to the formation of cD galaxies in clusters of Bautz-Morgan (BM) class I (Bautz \& Morgan 1970), since the performed analysis is simply applicable namely to clusters with a single very bright monster. It is assumed that cD galaxies become brighter by a continuous proccess of accretion of satellite galaxies.

\section{The Data and Analysis}

\subsection{The Data}

For study the possible mechanism of formation of BCGs I used the Abell, Corwin \& Ollowin (ACO) (1989) clusters of BM class I which contain a single very bright galaxy. The performed inspection of images of these clusters showed that clusters A447, A1508, A1775, A2025 mentioned as of BM type I in ACO, do not contain a single very bright galaxy. Therefore, they were excluded from the compiled list. Also, I found that radial velocities of bright galaxies in A1631 and A1654 differ significantly from that of the corresponding clusters. So these clusters are not of BM type I. The compiled list consists of 53 BCGs of which 34 are classified in Struble \& Rood (1987) as cD, and 19 as B, F, etc. 

I determined absolute stellar magnitudes of BCGs using the K$_{s-total}$ apparent magnitudes from 2MASS (Jarrett et al. 2000) which are more appropriate for the study the BCGs, since K magnitude encompasses the light of the predominantly red population in early-type galaxies. Note that 2MASS magnitudes have problems (e.g. Bell et al. 2003) in detecting low surface brightness light, such as halos of cD galaxies (e.g. Schombert 1988). In addition, Lauer et al. (2007) demonstrated that 2MASS photometry is likely to underestimate the total light from BCGs. At the same time, Lauer et al. (2007) showed that 2MASS photometry is free from possible errors which may be caused by the sky background substraction and crowding. The most important inconsistency may be produced by the extrapolation scheme to generate "total magnitudes" (Jarret et al. 2000). Meanwhile, Lin \& Mohr (2004) used a correction scheme to extrapolate isophotal magnitudes to "total" magnitudes and showed that both schemes are consistent. Summing up we may assume that mentioned problems of 2MASS magnitudes may not introduce systematic errors in our study. Note, that the 2MASS magnitudes have been widely used in galactic studies (e.g., Temi, Brighenti, \& Mathews 2008; Courteau et al. 2007; Masters, Springob, \& Huchra, 2007, etc.).

The $M_k$ absolute magnitudes were deduced using the known redshifts of clusters  (Andernach et al. 2005) in which corresponding BCGs recide. If the redshift of the cluster was unknown the absolute magnitude was determined using the redshift of the galaxy itself given in the NED (NASA Extragalactic Database). For distance determinations $H_0=72$ km $s^{-1}$ Mpc$^{-1}$ was adopted. A correction for the Galactic extinction was introduced according to Schlegel, Finkbeiner \& Davis (1998) given in NED, and the K-correction according to Kochanek et al. (2001). 

The list of the used BCGs are presented in Table 1 in the consecutive columns of which the following data are given: column 1 - the Abell cluster designation; column 2 the morphological type of the BCG given in Struble \& Rood (1987); column 3 - the redshift of the cluster (by * is marked the redshift of the galaxy from NED), column 4 - the absolute magnitude $M_k$ of the galaxy, column 5 - the Abell number count $N_A$\footnote
{the Abell number count $N_A$ is the number of galaxies more luminous than $m_{3+2}$ mag, where $m_3$ is the apparent photored magnitude of the third most luminous cluster member, located within one Abell radius $R_c$ of the cluster center. $N_A$ values are taken from Struble \& Rood (1987) who adopted for $R_c$ the apparent cluster radius estimated by Leir \& van den Berg (1977) or otherwise estimated by Abell (1980) and converted to Leir \& van den Berg system.};
column 6 - the estimated number $N_m$ of merged galaxies  (determined for the volume limited sample) (see below); column 7 - the velocity dispersion $\sigma_v$ (Andernach et al. 2005; Struble \& Rood 1999). 

\begin{table}[]
\caption[]{The list of cDs in Abell clusters of BM type I.}
\tabcolsep 8 pt
\begin{tabular}{llcclll} \\
\hline
Abell & Type & z & $M_k$ & $N_A$ & $N_m$ & $\sigma_v$ \\
      &      &   &       &       &       &  km s$^{-1}$ \\
\hline
21    & B  & 0.0955  & -27.32 &  56  & 72  & 855   \\
22    & B  & 0.1410* & -27.36 & 141  & 74  &  -    \\
42    & B  & 0.1129* & -26.69 & 154  & 40  &  -    \\
77    & cD & 0.0717  & -26.64 &  50  & 37  &  -    \\
85A   & cD & 0.0558  & -26.84 &  59  &	-  & 947   \\
122   & C  & 0.1128* & -27.41 &  64  & 80  &  -    \\
136   & cD & 0.1569* & -27.36 &  99  & 74  &  -    \\
146   & cD & 0.1878* & -27.47 &  -   & -   &  -    \\
192   & cD & 0.1219* & -27.00 &  90  & 59  &  -    \\
214   & B  & 0.1597* & -27.41 &  71  & 78  &  -    \\
261   & cD & 0.0469* & -25.92 &  63  & 20  &  -    \\
360   & cD & 0.2205* & -27.73 & 107  & -   &  -    \\
401   & cD & 0.0735  & -27.06 &  90  & 56  & 1083  \\
496   & cD & 0.0326  & -26.06 &  50  & 23  & 673   \\
586   & B  & 0.1700*  & -26.66 & 190  & -   &  -    \\
690   & cD & 0.0788* & -27.19 &  52  & 64  &  -    \\
733   & cD & 0.1159* & -26.92 &  64  & 44  &  -    \\
734   & cD & 0.0719* & -26.01 &  71  & 21  &  -    \\
882   & B  & 0.1407* & -25.69 &  48  & 16  &  -    \\
1068  & cD & 0.1375* & -27.07 &  71  &  57 &  -    \\
1146  & cD & 0.1412  & -28.10 & 222  & 147 & 1019  \\
1177B & cD & 0.0316  & -26.00 &  32  &     &  261  \\
1277  & B: & 0.2431* & -27.65 &  62  &  -  &  -    \\
1413  & cD & 0.1409  & -27.68 & 196  & 100 & 770   \\
1468  & C: & 0.0872  & -26.04 &  50  &  20 & 883   \\
1576  & B  & 0.2790* & -28.22 & 158  &  -  &  -    \\
1597  & cD & 0.1102* & -26.51 &  54  &  34 &  -    \\
1602  & cD & 0.2410* & -28.20 &  59  &  -  &  -    \\
1738  & cD & 0.1173  & -27.47 &  85  &  80 & 546   \\
1795  & cD & 0.0625  & -26.63 & 115  &  38 & 782   \\
1839  & C  & 0.1450* & -26.43 &  63  &  32 &  -    \\
1954  & cD & 0.1810* & -28.45 & 120  &  -  &  -    \\
1991  & F  & 0.0589  & -26.32 &  60  &  29 & 665   \\
2029A & cD & 0.0761  & -27.58 &  82  &  -  &  -    \\
2107  & cD & 0.0416  & -26.38 &  51  &  30 & 613   \\
2124  & cD & 0.0667  & -26.80 &  50  &  44 & 787   \\
2199  & cD & 0.0311  & -26.50 &  88  &  32 & 714   \\
2271  & cD & 0.0576* & -26.34 &  35  &  29 & 460   \\
2283  & L: & 0.1830* & -28.15 &  65  &  -  &  -    \\
2364  & F  & 0.1473* & -26.24 &  72  &  59 &  -    \\
2397  & cD & 0.2240* & -28.16 & 146  &  -  &  -    \\
2416  & cD & 0.2133* & -27.64 &  57  &  -  &  -    \\
2420  & cD & 0.0852  & -27.10 &  88  &  58 &  712  \\
2456  & C  & 0.0766* & -25.75 &  50  &  17 &  -    \\
2521  & B  & 0.1340* & -26.36 & 103  &  30 &  -    \\
2533  & cD & 0.1114* & -27.33 &  59  &  72 &  -    \\
2577  & I  & 0.1251* & -26.34 &  73  &  29 &  -    \\
2579  & B  & 0.1117* & -26.73 &  66  &  40 &  -    \\
2589  & cD & 0.0419  & -26.18 &  40  &  25 &  797  \\
2666  & cD & 0.0281  & -25.99 &  34  &  21 &  646  \\
2631  & cD & 0.2730* & -27.25 & 136  &  -  &  -    \\
2667  & I  & 0.2300* & -26.86 & 165  &  -  &  -    \\
2694  & cD & 0.0958* & -27.24 & 132  &  67 &  -    \\
\hline
\end{tabular}
\end{table}

\subsection{Formation and Evolution of cD galaxies}

The distribution of $M_k$ absolute magnitudes of BCGs against redshift in clusters of BM class I is shown in Fig. 1. The dotted lines show the locus of absolute magnitudes of galaxies the K$_{s-total}$ apparent magnitudes of which are equal to $9\fmag0$ and $14\fmag0$ respectively. For drawing these lines the K-correction (Kochanek et al. 2001) to apparent magnitudes was applied. All observed cDs and BCGs located within these two lines represent a complete sample. The absence of BCGs to the right of the line K$_{s-total}=14\fmag0$ is caused by selection effect: fainter objects are not detected. Fig. 1 shows that the sample of 41 BCGs is volume limited up to $z=0.1622$. Objects marked by open circle are outside the limiting distance of the volume limited sample of BCGs. The further discussion is performed using the volume limited complete sample.

\begin{figure}[htb]
\includegraphics[width=10cm]{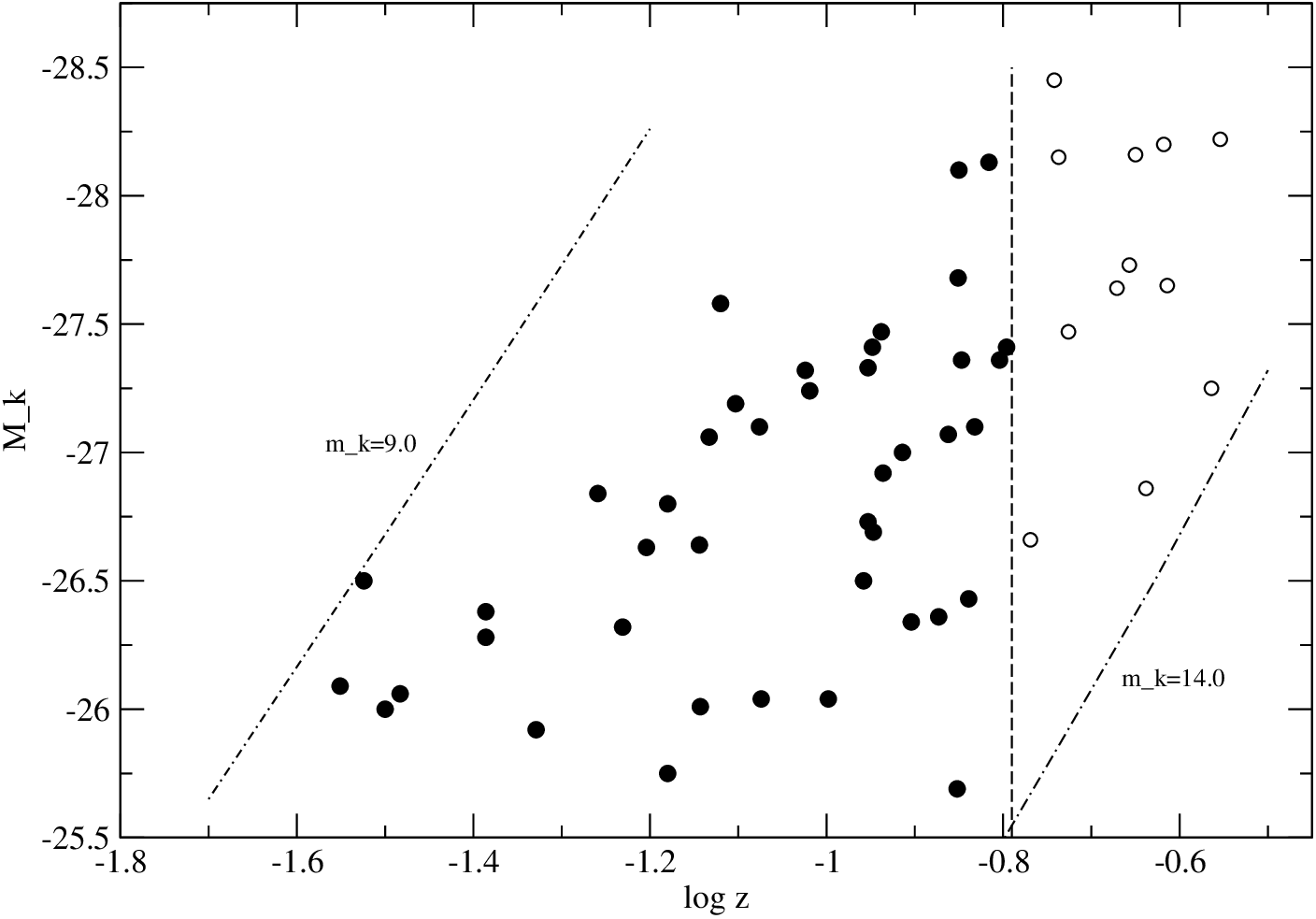}
\caption{The distribution of $M_k$ absolute magnitudes of BCGs (filled circles) in Abell clusters of BM class I against redshift. By open circles the galaxies at redshifts higher than $z=0.1662$ are shown.}
\label{fig1}
\end{figure}

In Fig. 2 the distribution by redshift of cDs and BCGs without halo classified as B, L, F, C in Struble \& Rood (1987) is presented. Fig. 2 shows the absence of galaxies without observed halo characteristic to cD galaxies at small redshifts and their higher number at high redshifts. One may assume that BCGs in all clusters of BM class I are mostly of cD type the supposed halo of which is not detectable at high distances. Remember that, as Sandage (1994) showed, the surface brightness $S_B$ of a galaxy depends very strongly on the distance: $S_B \sim 1/(1+z)^4$. Therefore, for study the formation of BCGs the common sample of cDs and other BCGs, named further on as cDs, is considered.

\begin{figure*}[htb]
\includegraphics[angle=0,width=6cm,keepaspectratio]{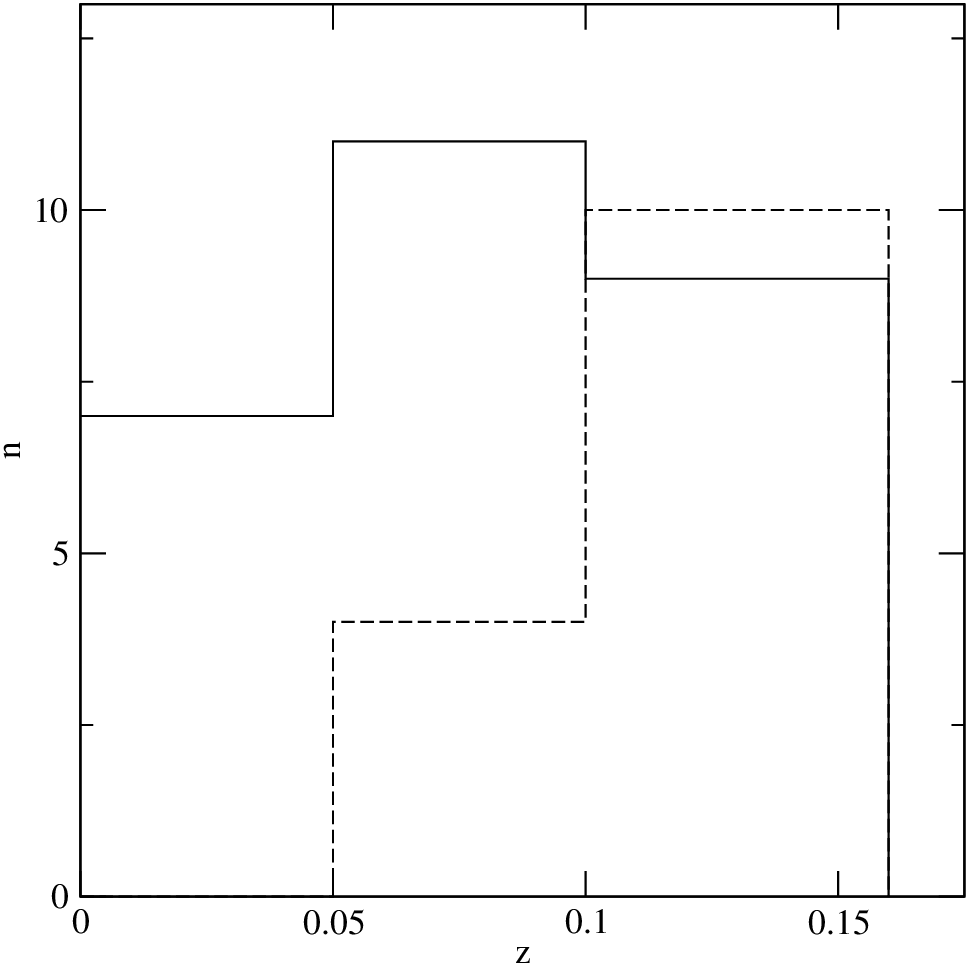}
\caption{The distribution of the number of cDs (solid line) and BCGs without halo (dashed line) by redshift.}
\label{fig2}
\end{figure*}

Fig. 1 demonstrates a peculiar distribution of cDs by redshift. The luminosity of the brightest cDs increases gradually with increase of redshift $z$. At the same time, contrary to Malmquist bias, the luminosity of the faintest cDs somewhat increases with decrease of redshift. I show below that such peculiar dependence of the luminosity on $z$ may be explained assuming that cDs are evolving due to cannibalism. 

If cDs are formed by cannibalism, they will be brighter in rich clusters, since the probability of encounter and merging obviously depend on the number of galaxies in a cluster. Indeed, the richer is a cluster the more members may be cannibalized by the forming cD in the potential well of the cluster, and the brighter will be the formed cD. The richness of a cluster is generally characterized by the Abell number count $N_A$. For estimation of the total number of galaxies within Abell radius we have to take into account the galaxies which already merged to the central monster. We may estimate the number $N_m$ of merged galaxies assuming that they are ordinary galaxies. For the absolute magnitude of the latter I used $M_{k(isol)}=-22\fmag68$ found by Tovmassian, Plionis \& Andernach (2004) for an isolated E/S0 or spiral galaxies. The number $N_m$ of merged galaxies is determined  by the formula $N_m=antilog[(M_{k}-M_{k(isol)}/2.5)$. If cDs are formed in the result of cannibalism, the faintest cD with $M_k\approx-26^m$ was formed by merging of about 21 galaxies with a mean brightness of an isolated E/S0 galaxy. The brightest cDs with $M_k$ equal to about $-28^m$ are formed then by merging of about 130 ordinary E/S0 or spiral galaxies. Adding the number $N_m$ of merged galaxies to the Abell number count $N_A$ we estimate the initial number $N_i$ of galaxies within one Abell radius before the start of merging. In total the initial number of galaxies in corresponding clusters was determined for 38 cDs of the volume limited complete sample. The number of merged galaxies and consequently the initial number of the cluster members was not determined for cDs located in clusters A85, A1177 and A2029 which according to Andernach et al. (2005) consist of separate projected clusters the Abell number count of which is thus uncertain.

Fig. 3 shows that there is a very strong corelation between the initial number $N_i$ of galaxies and the absolute magnitude $M_k$ of the formed cD expected in the case of formation of cDs by  cannibalism. Note, however, that $N_i-M_k$ correlation may be expected also when cDs are formed by rapid merging of galaxies during cluster collapse. 

\begin{figure*}[htb]
\includegraphics[angle=0,width=6cm,keepaspectratio]{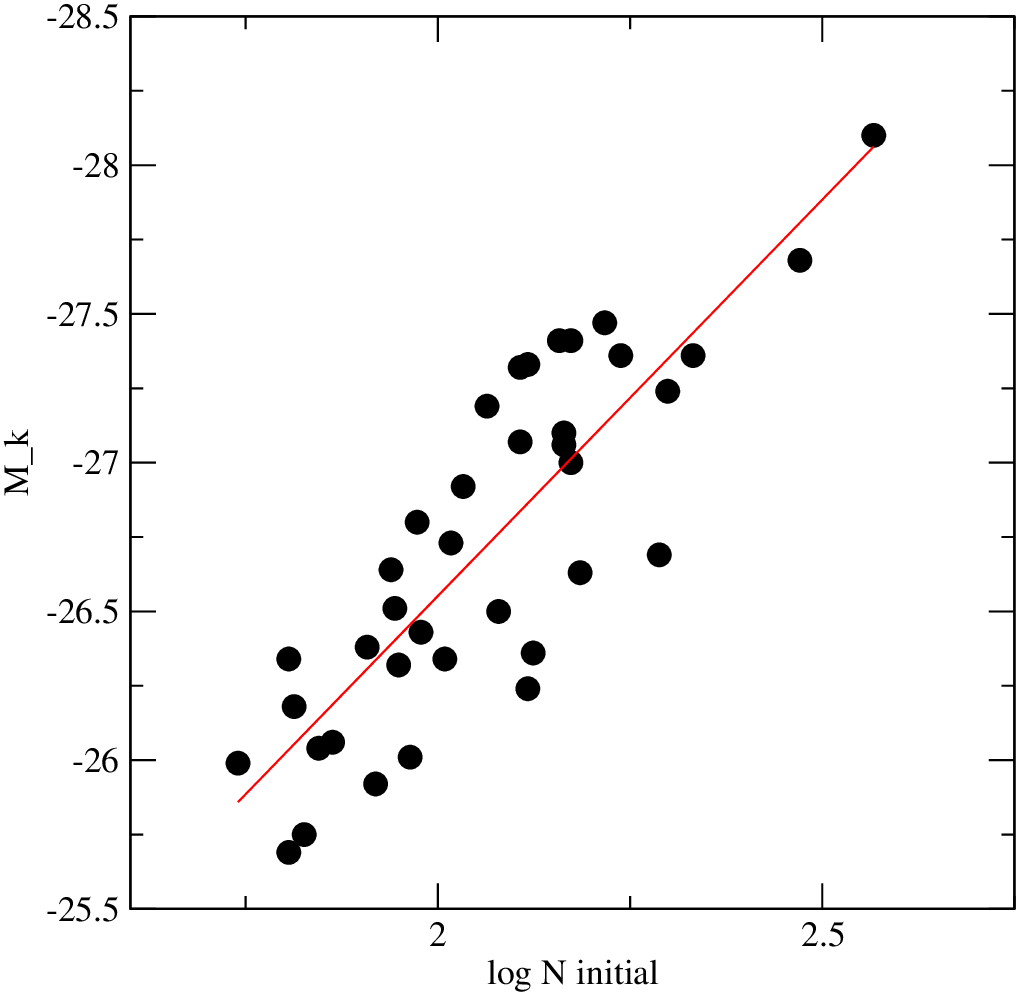}
\caption{The dependence of the absolute magnitude $M_k$ of a cD galaxy on initial number $N_i$ of galaxies in cluster.}
\label{fig3}
\end{figure*}

During permanent process of cannibalism the number $N_m$ of cannibalized galaxies by the cD galaxy obviously will increase in time. Therefore, one may expect that $N_m$ of galaxies cannibalized by older cDs in nearby space will be higher than cannibilazed by distant, younger ones. However, the supposed increase of the number of merged galaxies will be strongly masked by the following. As we saw, the brightness of the cD strongly depends on the initial number $N_i$ of galaxies in parent cluster (Fig. 3). On the other hand, the richness of a cluster depends on the distance. It is known that the farther is the cluster, the richer it is (Abell 1958). The richness-distance dependence of the used sample of cDs is presented in Fig. 4. Note that the richness of the rich clusters sharply decreases with decrease of distance, while for the poorest clusters this decrease is very slow. Hence, the farther is the cluster, generally the richer it is, and the brighter will be the cD galaxy formed in it. 

In order to overcome the difficulty to detect possible evolution of cDs, I compared the mean relative number $N_r=N_m/N_i$ of merged galaxies in small groups of cDs with small range of absolute magnitudes $M_k$. The sample of 38 cDs with estimated $N_r$ ratio was ordered by increasing $M_k$, and were divided into groups consisting of six galaxies each. Then the faintest galaxy from the first group was deleted and new groups of six galaxies were composed. This procedure was repeated five times. At the end the list of 33 independent groups consisting each of 6 galaxies with small range of absolute magnitudes was compiled. Replacement of one galaxy in a group of six galaxies resulted in different brightness and redshift limits, and also the different median redshift of the group. Therefore, all sets of six galaxies in the compiled six samples are independent. Then the mean relative numbers $N_r$ and corresponding dispersion for three nearby and three far galaxies in each group were determined.  The results on the composed 33 independent groups are presented in Table 2. The highest difference $\Delta M_k$ of absolute magnitudes of galaxies in each set of groups is also shown.

\begin{figure*}[htb]
\includegraphics[angle=0,width=6cm,keepaspectratio]{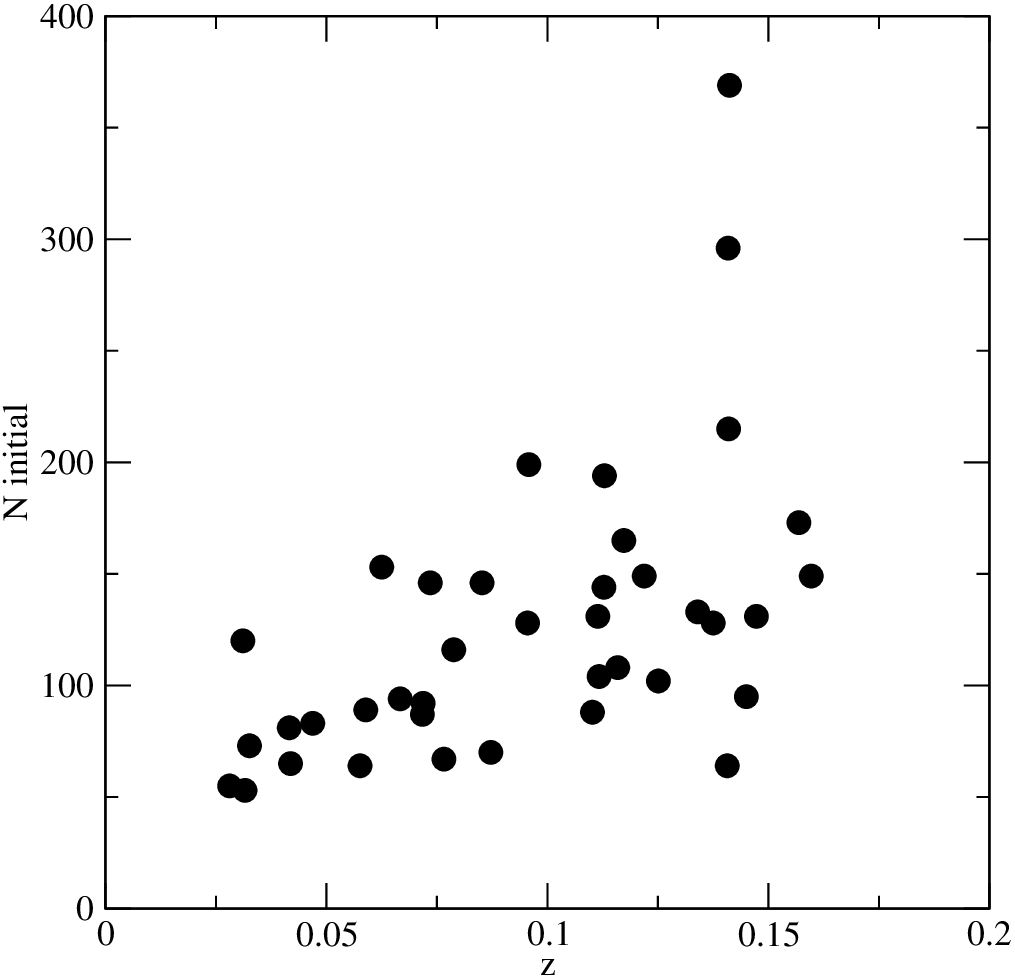}
\caption{The dependence of the initial number $N_i$ of galaxies in cluster on redshift.}
\label{fig4}
\end{figure*}

Inspection of Table 2 shows that in one group (group 4 in sample 4) the mean ratio $N_r$ for groups of nearby and far galaxies is the same, in seven groups (groups 2, 3, 4 in sample 1, group 3 in sample 2, groups 3 and 4 in sample 3, and group 3 in sample 4) the mean ratio $N_r$ is high for far galaxies. In the other 25 groups the mean ratio $N_r$ is higher for nearby galaxies. Though the difference between the mean $N_r$ of nearby and far galaxies in some groups is not high in comparison to corresponding dispersion, the probability that in 25 out of 33 groups the mean $N_r$ ratio for nearby galaxies will be by chance higher than that of for far galaxies is extremly small, $P=6\cdot10^{-7}$. Note that $N_A$ for far clusters may be underestimated. Therefore, the estimated ratio $N_r$ for far clusters may in fact be even smaller. This shows with certainty that nearby cDs in each group, i.e. the older ones cannibalized more cluster members than younger cDs in more distant clusters with about the same richness. Hence, the performed analysis shows that in spite of the increase of the richness of clusters with redshift (Fig. 4) and the corresponding increase of brightness with cluster richness (Fig. 3), the number of merged galaxies in the samples of cDs with about the same brightness increases with decrease of redshift. It means that cDs in clusters of BM type I grow in time. They ``eat`` more cluster members during their life. The obtained results are in favor of the suggestion that cDs are formed by the mechanism of cannibalism, and are {\it evolving}. 

In the result of cannibalization of cluster members cD galaxies become brighter in time and move up on the diagram of Fig. 1. This process for cDs in rich clusters will last relatively long time for two reasons. At first, the rich and massive clusters have high velocity dispersion $\sigma_v$ (see Fig. 5), and for this reason the merging process in them will be slow. And at second, for the large stock ($N_i$) of candidate galaxies for merging. Thus, cDs in rich clusters slowely move up forming the upper envelope of the distribution of the cD brightness against $z$ (Fig. 1). In poor clusters with small $\sigma_v$ the rate of merging is obviously higher than in rich clusters. At the same time, the resources of galaxies for merging are small. Therefore, the process of cannibalism in poor clusters may terminate faster. In relatively short time there will be very little amount of galaxies in poor clusters to be "eaten" by the forming cD, and they shortly will reach their possible, not high brightness. And the older is the formed cD the more galaxies it cannibalized, and the brighter it will be. This is demonstrated by the lower envelope of the distribution in Fig. 1. Note that the initial number $N_i$ of the poorest groups gradually increases with redshift (Fig. 4). Therefore, since the brightness of the formed cD depends on the richness of the cluster, the opposite trend should be observed, i.e. the brightness of cDs in poorest clusters will increase with redshift. This obviously decreases the found trend of increase of the brightness of the poorest cDs with decrease of redsift which in reality will be stronger than the observed one. Thus, the suggested scenario explains the upper and lower envelopes of distribution of $M_k$ of cDs against $z$ demonstrated in Fig. 1. Described scenario of evolution of cD galaxies explains the peculiar luminosity function of cD galaxies presented in Lopes-Cruz, Tovmassian \& Mendes de Oliveira (2008).

\begin{figure*}[htb]
\includegraphics[angle=0,width=6cm,keepaspectratio]{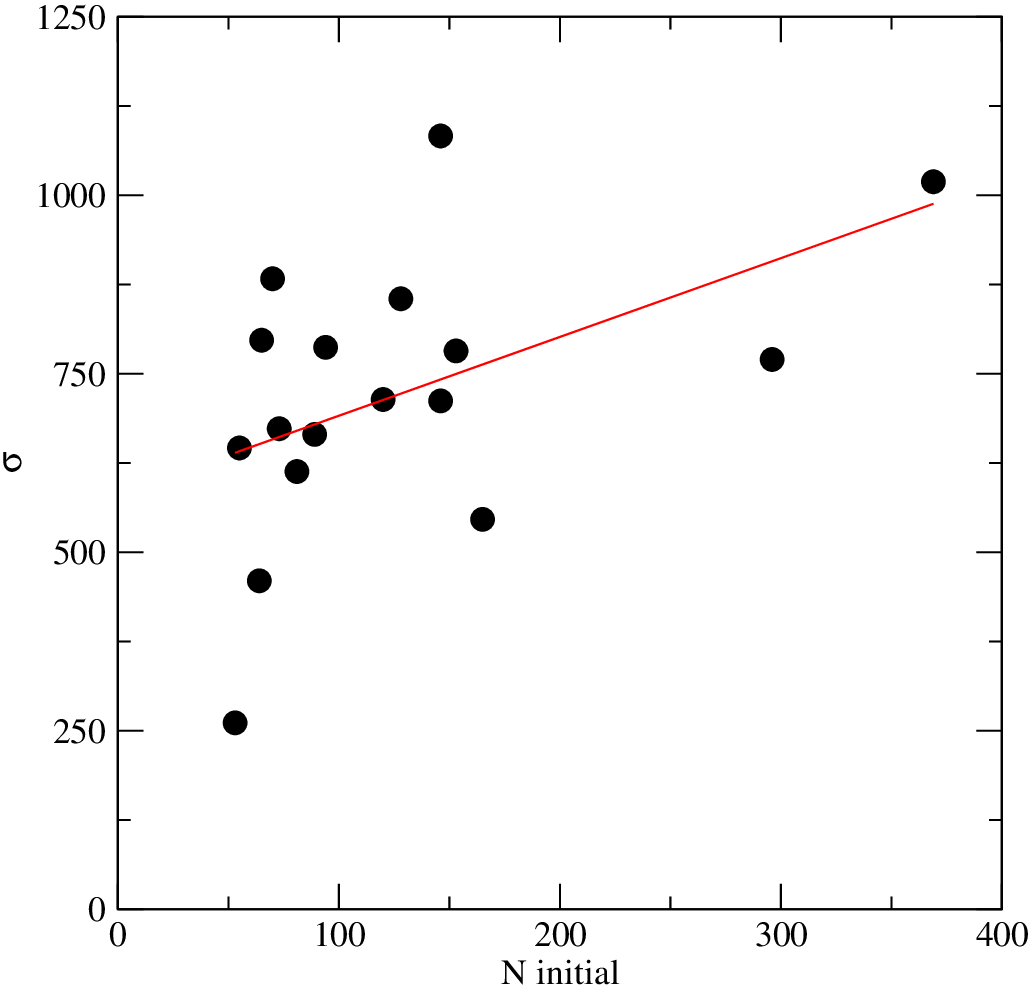}
\caption{The dependence of the velocity dispersion $\sigma_v$ on the initial number $N_i$ of galaxies in cluster.}
\label{fig5}
\end{figure*}

{\scriptsize
\begin{table}[htb]
\caption[]{Comparison of the mean ratio $N_m/N_i$ for nearby and far galaxies in bins with small range of absolute magnitudes $M_k$.} 
\begin{tabular}{ccccccc} \\ 
\hline
Group & group 1 & group 2 & group 3 & group 4 & group 5 & group 6  \\ 
\hline
\hline
  &  & & Sample 1 &   \\
\hline
Nearby & $0.283\pm0.068$ & $0.342\pm0.031$ & $0.295\pm0.054$ & $0.336\pm0.092$ & $0.444\pm0.076$ & $0.530\pm0.073$ \\
Far    & $0.263\pm0.019$ & $0.396\pm0.079$ & $0.316\pm0.067$ & $0.423\pm0.037$ & $0.436\pm0.078$ & $0.485\pm0.032$  \\
$\Delta M_k$ & $0\fmag35$ & $0\fmag28$     & $0\fmag27$      & $0\fmag36$      & $0\fmag26$      & $0\fmag14$      \\
\hline
  &  &  & Sample 2 &  \\
\hline
Nearby & $0.313\pm0.058$ & $0.388\pm0.052$ & $0.295\pm0.054$ & $0.410\pm0.041$ & $0.492\pm0.068$ & $0.460\pm0.091$ \\
Far    & $0.256\pm0.024$ & $0.320\pm0.095$ & $0.383\pm0.036$ & $0.336\pm0.092$ & $0.444\pm0.087$ & $0.432\pm0.073$ \\
$\Delta M_k$ & $0\fmag31$ & $0\fmag18$      & $0\fmag26$      & $0\fmag37$      & $0\fmag26$      & $0\fmag32$      \\
\hline
  &  & & Sample 3 & & \\
\hline
Nearby & $0.361\pm0.032$ & $0.383\pm0.053$ & $0.310\pm0.076$ & $0.410\pm0.041$ & $0.492\pm0.068$ & $0.460\pm0.091$ \\
Far    & $0.252\pm0.025$ & $0.320\pm0.095$ & $0.313\pm0.079$ & $0.416\pm0.021$ & $0.410\pm0.099$ & $0.450\pm0.033$ \\
$\Delta M_k$ & $0\fmag26$ & $0\fmag12$      & $0\fmag26$      & $0\fmag34$      & $0\fmag26$      & $0\fmag74$     \\

  &  & & Sample 4 & & \\
\hline
Nearby & $0.361\pm0.032$ & $0.383\pm0.053$ & $0.313\pm0.079$ & $0.416\pm0.037$ & $0.472\pm0.096$ & - \\
Far    & $0.321\pm0.094$ & $0.282\pm0.045$ & $0.323\pm0.083$ & $0.416\pm0.021$ & $0.441\pm0.085$ & - \\
$\Delta M_k$ & $0\fmag25$ & $0\fmag11$      & $0\fmag26$      & $0\fmag34$      & $0\fmag17$   & -   \\
\hline
  &  & & Sample 5 & & \\
\hline
Nearby & $0.342\pm0.031$ & $0.363\pm0.074$ & $0.380\pm0.095$ & $0.444\pm0.076$ & $0.471\pm0.095$ & - \\
Far    & $0.321\pm0.094$ & $0.282\pm0.045$ & $0.323\pm0.083$ & $0.416\pm0.021$ & $0.443\pm0.087$ & - \\
$\Delta M_k$ & $0\fmag31$ & $0\fmag16$      & $0\fmag29$      & $0\fmag27$      & $0\fmag17$      & - \\
\hline
  &  & & Sample 6 & & \\
\hline
Nearby & $0.384\pm0.056$ & $0.341\pm0.053$ & $0.380\pm0.095$ & $0.444\pm0.076$ & $0.544\pm0.012$ & - \\
Far    & $0.354\pm0.070$ & $0.282\pm0.045$ & $0.330\pm0.088$ & $0.393\pm0.044$ & $0.432\pm0.073$ & - \\
$\Delta M_k$ & $0\fmag30$ & $0\fmag17$      & $0\fmag29$      & $0\fmag24$      & $0\fmag09$     & - \\  
\hline
\end{tabular}
\end{table}
}

\subsection{Fossil groups.}

Fossil groups fit well with the scenario of evolution of cD galaxies by cannibalism. Vikhlinin et al. (1999) and Mulchaey \& Zabludoff (1999) expressed an idea that fossil groups could be formed by complete merging of galaxies within once normal clusters. In Fig. 6 we present distributions $M_k-log~z$ for the complete sample of cDs in clusters of BM class I, and for the brightest galaxies in fossil groups for the same distance limit. We used the list of the fossils compiled by Santos, Mendes de Oliveira \& Sodre (2007). Fig. 6 shows that almost all fossil groups are among faint cDs. It means that the initial number of galaxies in the primordial fossil clusters was small. As generally in poor clusters, after merging of some number of galaxies there will be almost no more galaxies left for merging with the bright galaxy, and the cluster will become a fossil group.

\begin{figure*}[htb]
\includegraphics[width=100mm]{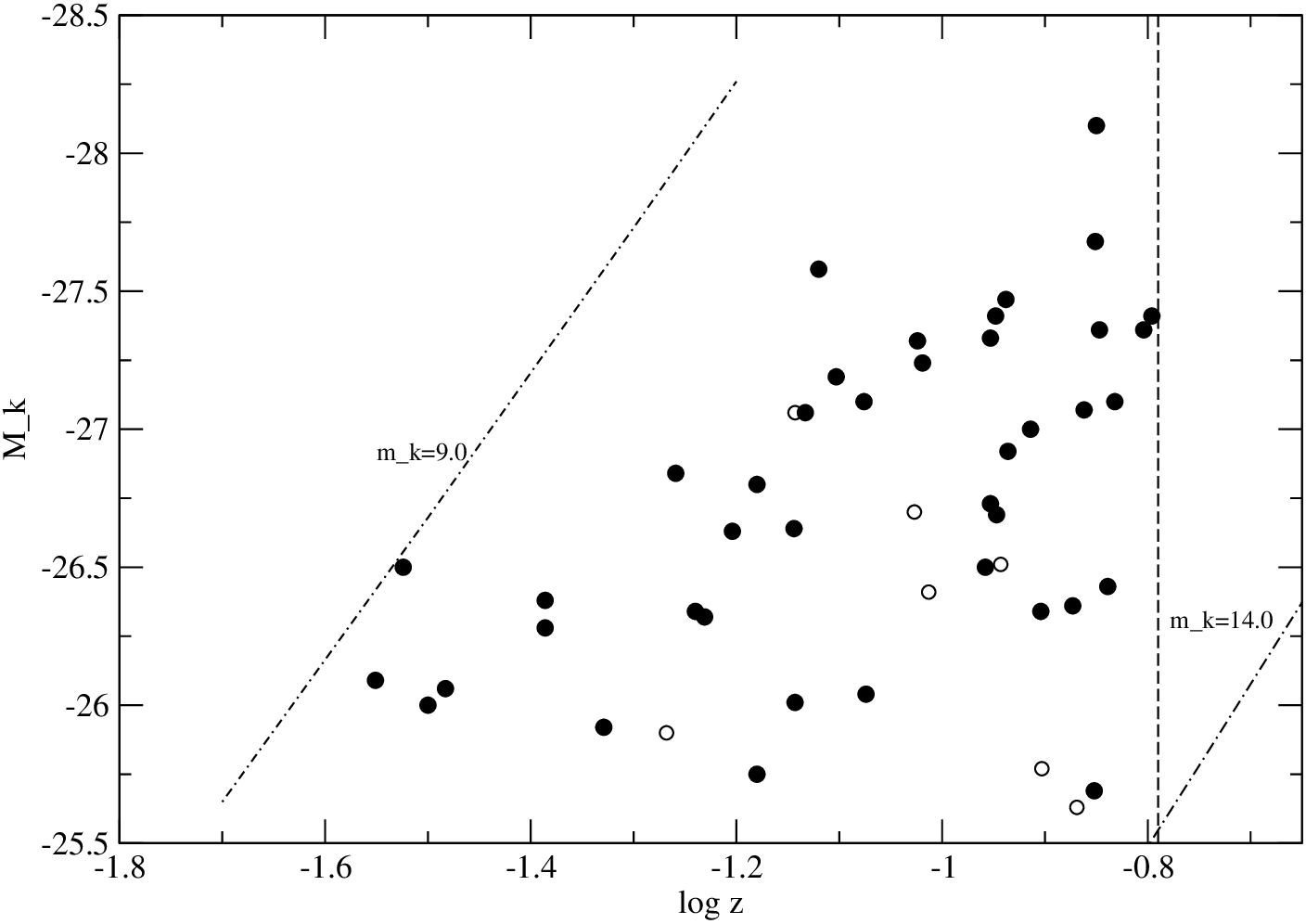}
\caption{The distribution of $M_k-log~z$ of the volume limited complete sample of cDs in BM class I clusters (filled circles) and the brightest galaxies in fossil froups (open circles).}
\label{fig6}
\end{figure*}

\section{Conclusions}

I present evidences that cDs in clusters of galaxies of BM class I are formed by the mechanism of cannibalism. The process of cannibalism in a cluster is a permanent one. In rich clusters more galaxies may merge in time to the central monster, and the brightness of the latter may grow to very high value. The process of cannibalism in rich clusters is a relatively long lasting one for small rate of merging (high velocity dispersion) and large number of galaxies in the cluster. The cDs formed in rich clusters determine the upper envelope of the cD brightness distribution by redshift (Fig. 1). The cDs formed in poor clusters may be not very bright for small stock of galaxies. The process of cannibalism in poor clusters terminates faster, first for the high rate of merging (small velocity dispersion), and second, for the small number of galaxies for merging. 
Nearby, older cDs have ``eaten`` more cluster members in comparison to cDs in distant, younger clusters with about the same richness. Old cDs in nearby poor clusters already reached their possible high brightness. The older is the cD the brighter it is. This determines the peculiar lower envelope of the cD brightness distribution (Fig. 1). In the frames of this scenario fossil groups are those poor clusters in which the process of cannibalism already finished due to the small stock of galaxies in the system, and the brightness of the central monster practically reached its possible high value.

\begin{acknowledgements}
I am grateful to the anonymous referee whose critical comments forced me to make additional analysis and largely rewright the paper. I thank O.Lopes-Cruz for useful discussion and H.Andernach for presenting an updated compilation of Abell cluster redshifts and velocity dispersions. This research has made use of the NASA/IPAC Extragalactic Database (NED) which is operated by the Jet Propulsion Laboratory, California Institute of Technology, under contract with the National Aeronautics and Space Administration.
\end{acknowledgements}

\end{document}